\documentclass[aps,twocolumn,superscriptaddress]{revtex4-1}
\usepackage{amsmath,amssymb}
\usepackage{graphics,graphicx}
\usepackage{dcolumn,bm}
\usepackage{psfrag}
\usepackage{color}
\usepackage{multirow}
\usepackage{dcolumn,bm}
\usepackage{bm}

\newcommand{\cu}
{\affiliation{Department of Physics, University of Calcutta,
92 Acharya Prafulla Chandra Road, Kolkata 700009, India.}}

\newcommand{\victo}
{\affiliation{Department of Physics, Victoria Institution (College),
78B Acharya Prafulla Chandra Road, Kolkata 700009, India.}}

\begin{document}


\title{Virtual walks and phase transitions in two dimensional
BChS model with extreme switches}

\author{Kathakali Biswas}
\email{kathabiswas2012@gmail.com}
 \victo
 \cu

\author{Parongama Sen}%
 \email{parongama@gmail.com}
\cu




\begin{abstract}
We have studied a walk in a one-dimensional virtual space corresponding to an extended version of the three-state BChS model of opinion formation, originally proposed in Physica A {\bf 391}, 3257 (2012), in which the agents are located on a two dimensional lattice. The opinions are designated by the values $\pm1$ and zero. Here we also consider switches between the extreme states $\pm 1$. The model involves two noise parameters representing the fraction of negative interactions $p$ and the probability of extreme switch denoted by  $q$. The study shows that the nature of the walks changes drastically as the noise parameters exceed certain threshold values. The order-disorder phase transitions are independently obtained using the finite size scaling method showing that these threshold values are indeed consistent with those values of the parameters where a phase transition exists. The criticality is 
found to be Ising-like even when extreme switches are allowed.   A new critical exponent associated with the probability distribution of the displacement is also obtained independent of the values of the critical parameters. The nature of the walks is compared to similar virtual walks studied earlier.
\end{abstract}

\maketitle


\section{Introduction}

In the past few decades, extensive research has been made to study the problem of opinion formation in a society, using the tools of statistical physics \cite{sociophysics,soc_rmp,galam_book}. Several models of opinion dynamics have been proposed in the past. One class of models, namely, the kinetic exchange (KE) models  involves an  interaction between two  agents at any instant \cite{kem_eco_sco}. A particular KE model,  popularly called the BChS model, incorporates negative interactions as well \cite{BCS}. This model yields an interesting phase transition. It has also been possible to obtain some results from this model that correspond to realistic scenarios \cite{us_elec_1,kbps1,brexit}. 

It is often convenient to study a model in statistical physics by mapping it into another model and studying the latter. In recent times, various models of statistical physics involving dynamics of spins, opinions, and financial status have been mapped to walks in a virtual space 
\cite{godre1,newman,baldasari,godre2,godre3,godre4,privman,derrida1,derrida2,derri2,howard,sbprps,pratik,surojit1,sanchari_walk,sanchari}. In these mappings, a walker is associated with each spin/agent. The position of the walker is updated according to the state of the spin/agent it is associated with, following 
either a Markovian or a non-Markovian dynamics. The nature of the walks usually undergoes
a change at the phase transition points, if any. This is
indicated by the existence of a Gaussian distribution of
the displacements of the walkers above the critical point
while it has a double peaked non-Gaussian behavior be-
low it. New critical exponents associated with the width of the distribution have been seen to exist as well. It is possible to conjecture whether the ordered state is an absorbing state by studying the scaling behavior of this distribution. 
An interesting crossover behavior involving diverging timescales, not obtained directly from the model, has been observed previously also \cite{pratik,surojit1}.

In this paper, we have studied a KE model  of opinion formation, an extended version of the so-called BChS model. In this opinion dynamics model, the agents are embedded on a two-dimensional (2$d$) lattice while the walks are defined in a virtual one-dimensional (1$d$) space.

The BChS model is taken with three opinion states denoted by $\pm 1 $ and $ 0$. The interactions here can be negative with probability $p$. In the original version \cite{BCS}, switches between the extreme states (i.e. $+1$ and $-1$) were not possible according to the dynamical rules. In some recent works, the present authors introduced the possibility of extreme switches occurring with probability $q$ \cite{kbps2,kbps3}. The mean field case was studied to show that a phase diagram can be obtained in the $p-q$ plane showing the presence of ordered and disordered regions \cite{kbps3}.

The critical behavior of the BChS model on two-dimensional square lattices for $q=0$ has been studied earlier using numerical simulations \cite{sudip,croki}. It was
found that the model belongs to the two-dimensional Ising criticality class. In this paper, we have simulated the model in 2$d$ with extreme switches, i.e., $q\neq 0$, and obtained 
the corresponding walk from which the phase diagram can be estimated. 
In addition, for comparison, a few phase transition points have been found
directly by analyzing the relevant physical quantities and using finite size scaling.
From the results obtained in the mean-field case \cite{kbps3}, we expect that $q$ will effectively impart an additional noise. It is also interesting to see whether the critical behavior is affected by $q$; in the mean-field case, it is not. 


The walks are also analyzed quantitatively by fitting appropriate curves to the distributions and studying the fitting parameters. This leads to some further
understanding of the system both quantitatively and qualitatively.

In section II, we describe the model and the methods used. The results are presented in section III and in the last section, discussions and conclusive statements have been made. 

\section{Model and Method}

\subsection{Simulating BChS model on a two dimensional  lattice with the parameters $p$ and $q$}
We have simulated an agent-based BChS model where agents are located on the sites of a 2$d$ square lattice with $N$ sites  ($N=L\times L$) having periodic boundary conditions. $o_i$ is considered as the opinion of the $i$th agent. Each opinion ($o_i=\pm1,0$)  is updated upon interaction with one of the nearest neighbors denoted by $k$, with opinion $o_k$, following the expression,

\begin{equation}
o_i(t+1)=o_i(t)+ \mu o_k(t).
\label{main1}
\end{equation}
 Here $\mu$ is the interaction parameter;  $|\mu|=1,2$ with probability $(1-q)$ and $q$ respectively. $\mu$ can be negative with probability $p$ for both values of $|\mu|$.
If after an interaction the opinion exceeds $1$ or becomes less than $-1$, it is adjusted to $\pm 1$ respectively. 

In the simulations, a homogeneous disordered state  is  considered as the initial 
configuration, i.e.,  equal proportion of the population   has opinion  $\pm 1$ and zero. 

\subsection{Mapping the BChS model into a Virtual Walk}

Mapping of the opinion dynamics model to a virtual walk is done by associating a virtual walker with each agent on the 2$d$ lattice. The virtual walk takes place in one dimension. Hence we have a scenario of $N$ walkers performing walks on a one dimensional lattice which is strictly speaking, unbounded. The walks are not independent as they are generated from the interaction of the agents.

In  the mapping scheme, the initial position of a walker is taken to be $0$. The  walks are implemented  according to the opinions of the agents which are updated asynchronously. 
 The total distance travelled from the starting point (at $t=0$) by the $i$th walker at the  $(t+1)$th Monte Carlo (MC) step is  $X_i(t+1)$, given by 
\begin{equation}
X_i(t+1)=X_i(t) + o_i(t+1).
\label{main2}
\end{equation}
Only Markovian walks have been considered in the present work.

We have simulated the system on $L \times L$ lattices with various values of $L$. For the virtual walk, 
the results for $L=64$ (maximum size simulated) only are presented. 
The walk is incremented by the value of the opinion after 
the completion of one Monte-Carlo step (MCS). One agent is selected randomly in time $\frac{1}{N}$ and allowed to interact with any of her four nearest neighbour after which her opinion is updated;  one MCS step comprises of $N$ such updates. 

\subsection{Finite size scaling analysis}

For a system manifesting a continuous phase transition driven by a certain parameter, the critical point and the exponents can
be obtained using finite size scaling method in a numerical simulation. Consider a physical  quantity $\Phi$, which either goes to zero or diverges in the 
manner $\Phi \propto \epsilon ^\phi$, where  $\epsilon$ is the small deviation from the critical point. 
In a finite system of size $L$, $\Phi$ can be expressed as    
\[
\Phi = L^{-\phi/\nu} f(\epsilon L^{1/\nu}),
\]
where $\nu$ is the correlation length exponent.
$f(z)$ is expected to vary as $z^\phi$ as $z \to \infty$ such that one recovers the behaviour $\Phi \propto \epsilon^\phi$ for $L \to \infty$.

A data collapse can be obtained (i.e., all data for different system 
sizes fall on the same curve) when properly rescaled quantities are plotted with accurate choices of the values 
of the critical point and the exponents.

\section{Results}
We have determined the probability $S(X,t)$ associated with the distance $X$ covered by a walker at a specific time $t$ by averaging over many configurations. In the Ising and voter model in 2$d$ and also in the mean-field KE model \cite{godre4,pratik,surojit1}, the nature of $S(X,t)$ has been observed to change at the phase transition points. 
Here also we have observed a similar change beyond threshold values of the parameters $p,q$.
We investigated the characteristics of the distribution both below and above the threshold points. Additionally, by performing data collapse analysis of the scaled data from the simulations conducted on 2$d$ lattices, we identified the critical points and estimated the critical exponents. 
The latter study has been made at larger times such that the system definitely reaches equilibrium.

\subsection{ Features studied from the Virtual walk}
\begin{figure}[h]
    \centering
    \includegraphics[width=8.5 cm]{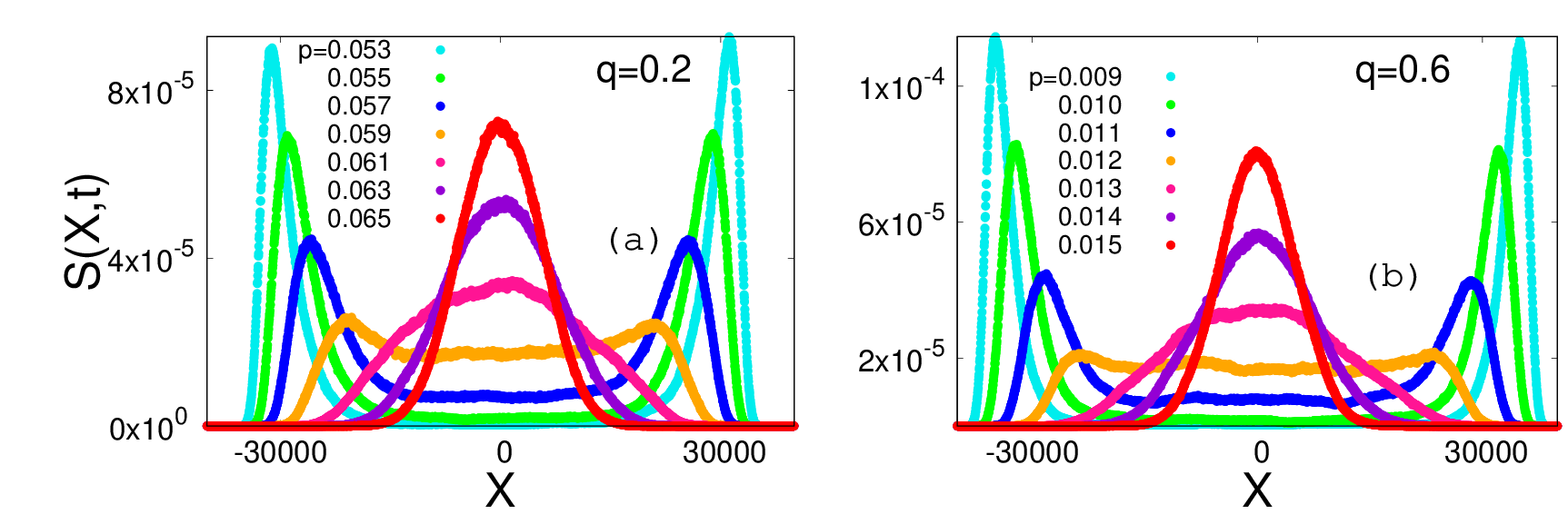}
    \caption{The probability distribution of $X$ at time step $t=60000$, when the model has reached a steady state, has been plotted for two different values of $q$ ($0.2$ in (a) and $0.6$ in (b)) for various $p$ values. These plots are presented for a system size of $N=64\times 64$. 
    \label{dist1}}
\end{figure}

In the $(p,q)$ parameter space, the threshold points, denoted as ($p_T, q_T$), have been determined using the virtual walk analysis. In general, we have kept $q=q_T$ fixed and varied $p$ to study the behavior of the walk. It is observed that when $p$ is less than $p_T$, the distribution  $S(X,t)$ exhibits a non-Gaussian, double-peaked behavior. Above the threshold point, the distribution assumes a centrally peaked Gaussian shape. The data are shown in Fig. \ref{dist1} for two different  $q_T$ values. 

\begin{figure}[h]
\centering
\includegraphics[width=7 cm]{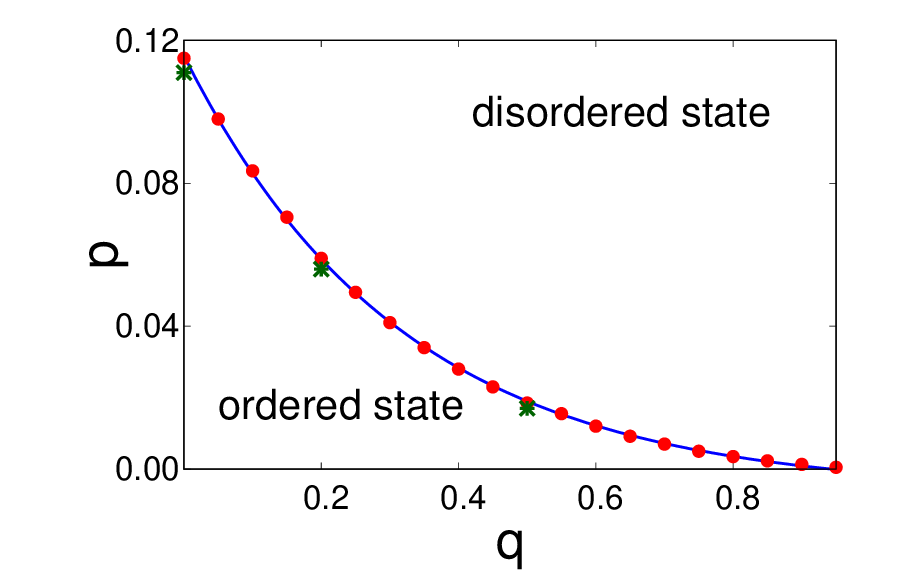}
\caption{The red dots in the $p,q$ parameter space represent threshold values separating the ordered and disordered phases. The phase boundary  can be fitted to the form  $p_T = C_0 \exp^{-\lambda q_T} - C_1$ with $\lambda=3.14957~ \pm 0.02697$, $C_1=0.00626~ \pm 0.00033$ and $C_0=0.12203~ \pm 0.00036$. Three critical points directly obtained from the simulation for $q=0, 0.2, 0.5$ using finite size scaling are also shown by the green points in the figure.}
\label{phase-dia}

\end{figure}

The determination of the threshold points $p_T$ has been carried out for several values of  $q$  within the range of $[0, 1)$, and as displayed in Fig. \ref{phase-dia}  these points can be fitted to the form
\begin{equation}
p_T \approx C_0 \exp^{-\lambda q_T} - C_1,
\label{boundary}
\end{equation}
where $\lambda\approx3.14$, $C_1\approx0.0063$ and $C_0\approx0.1220$. Specifically, one gets $p_T\rightarrow 0$ for $q_T=1$, exactly as in the case of the mean-field version of the present model \cite{kbps3}.

\begin{figure}[h]
\centering
\includegraphics[width=8.5 cm]{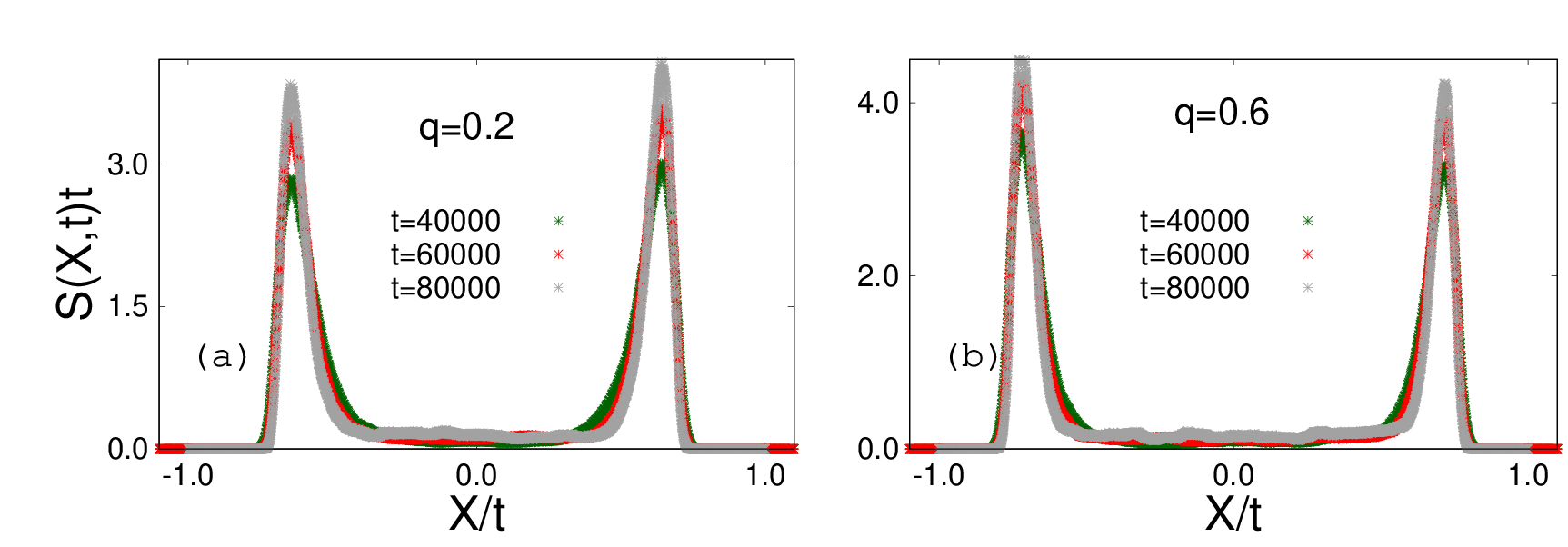}
\caption{The data collapse of $S(X,t)t$ using the scaling variable $X/t$ is illustrated for two different points: (a)  $q=0.2,~p=0.055$ 
and  (b) $q=0.6,~p=0.010$, both of which lie below the phase boundary (eq. \ref{boundary}).  }
\label{collapse1}
\end{figure}

\begin{figure}[h]
   \centering
    \includegraphics[width=8.5 cm]{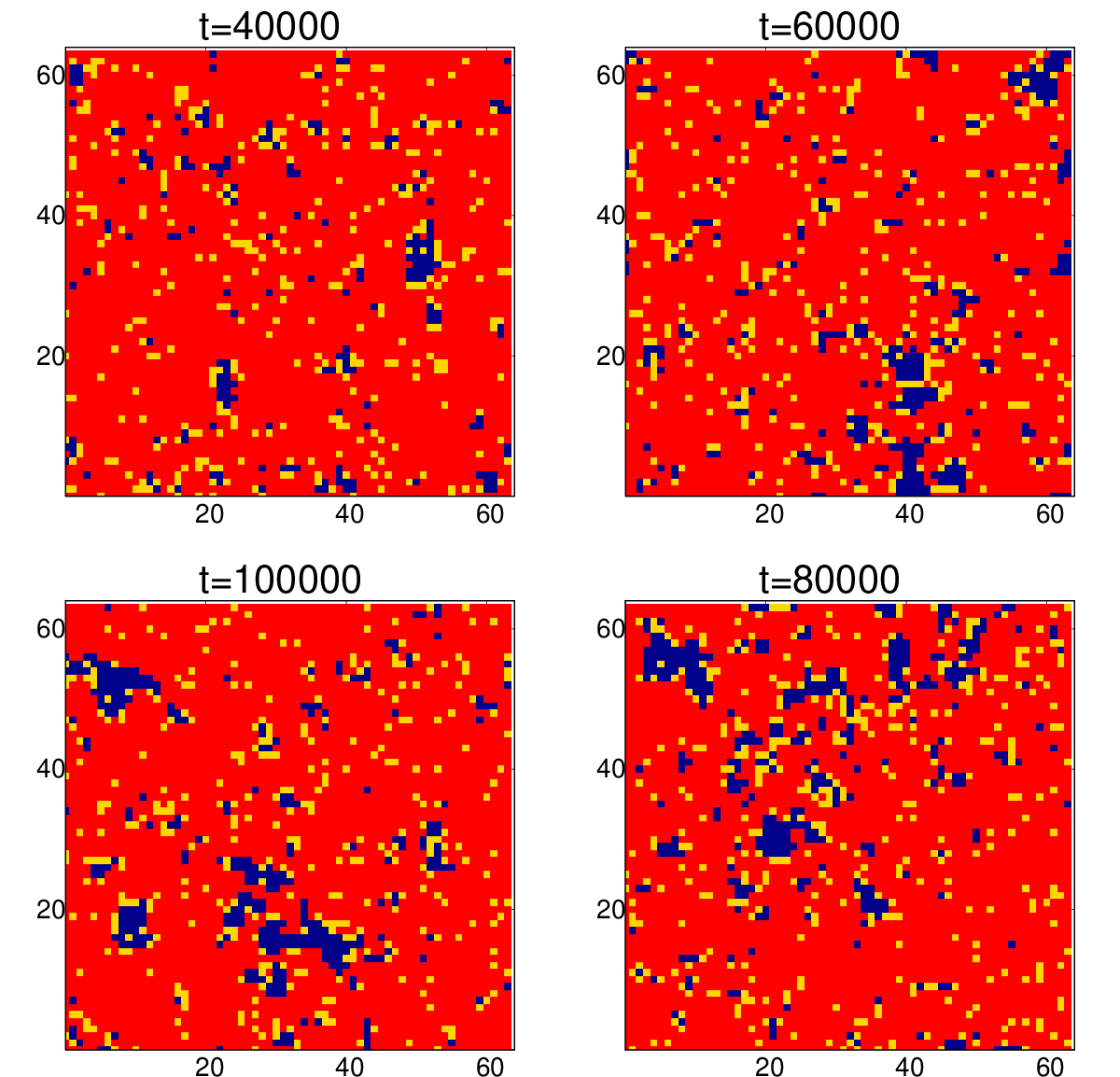}
    \caption{Snapshots of the 2$d$ lattice space at different time steps below the phase boundary. Red, blue, yellow dots denote $\pm 1$ and $0$ states respectively.
    \label{snapshot_below_pc}}
\end{figure}

To determine the nature of the walk, $S(X,t)$ is fitted to the scaling form 
\begin{equation}
S(X,t)\approx t ^{-\alpha}F(X/t^{\alpha}).
\label{main3}
\end{equation}
When $p<p_T$, an approximate data collapse has been obtained with $\alpha=1.0$ (shown in Fig. \ref{collapse1}). $\alpha = 1$ implies a ballistic walk. For an absorbing phase, a perfectly ballistic behavior is expected at late times. However such is not the case here. 
This indicates that for $ p < p_T$, where the ordered phase exists as is concluded from the nature of $S(X,t)$, we have an active state. Indeed,
the snapshots shown in Fig. \ref{snapshot_below_pc} 
 taken at different time show that the system is still active.
It is understandable why an active phase is present;   in the presence of noise, some agents will change opinion even at large times. In previous work, only for the noise-less case, it was shown that an absorbing phase is reached, although for a considerable number of configurations, it becomes a very slow process \cite{brexit}. We will come back to this point in the next section.

\begin{figure}[h]
\centering
\includegraphics[width=8.5 cm]{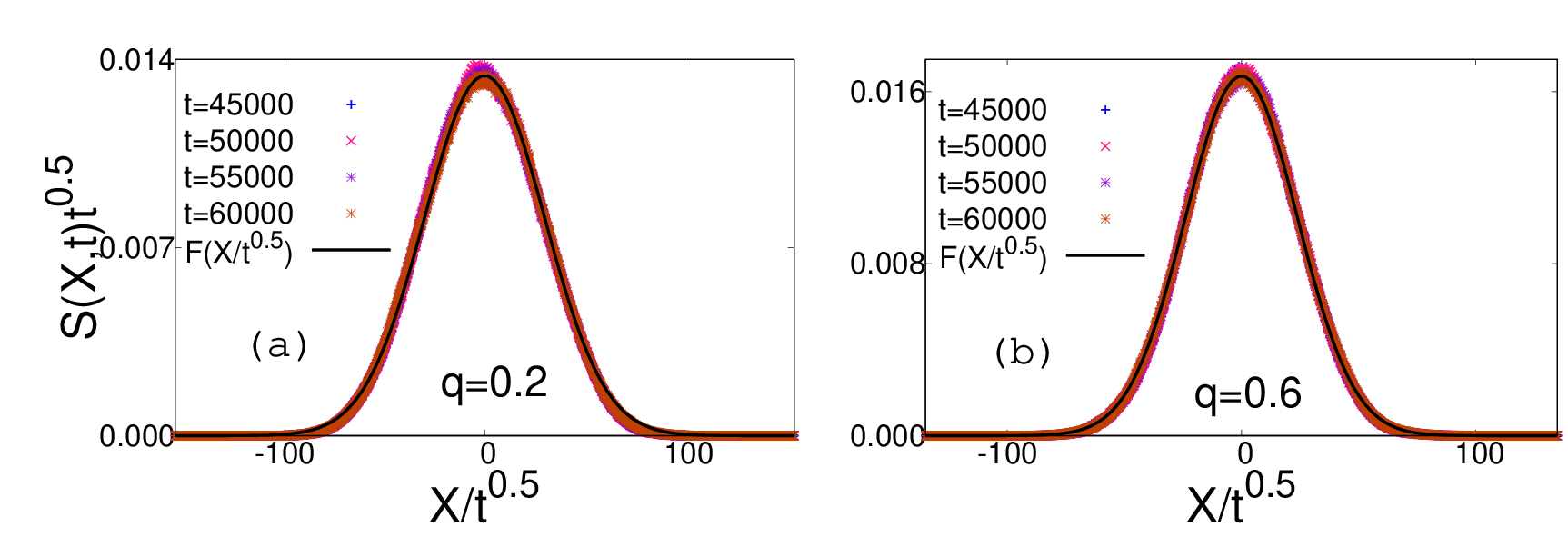}
\caption{ The data collapse of $S(X,t)t^{0.5}$ using the scaling variable $X/t^{0.5}$ is illustrated for two  cases (a) $q=0.2,p=0.064$ and (b) $q=0.6,p=0.015$, both of which lie above the phase boundary (i.e $p < p_T$ in each case).  Both sets of data can be fit using a Gaussian function denoted by $F(z)$ in the figures, where $z=\frac{X}{t^{1/2}}$.  }
\label{collapse2}

\end{figure}

\begin{figure}[h]
\centering
\includegraphics[width=7 cm,height=5cm]{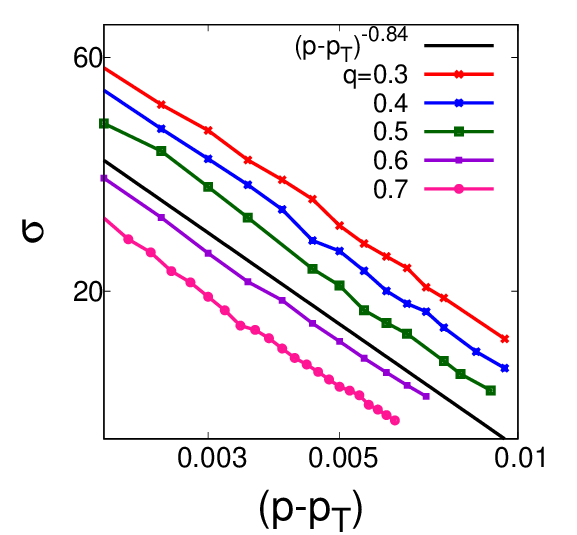}
	\caption{ The variation of the width $\sigma$ of the Gaussian functions as a function of $p$ above $p_T$ is shown in this figure for several $q$ values. The figure shows that the nature of the curves is compatible with a variation $\sigma \sim (p-p_T)^{-\delta}$ with  $\delta \approx 0.84$.
}
\label{width}
\end{figure}

\begin{figure}[h]
    \centering
    \includegraphics[width=8.5 cm]{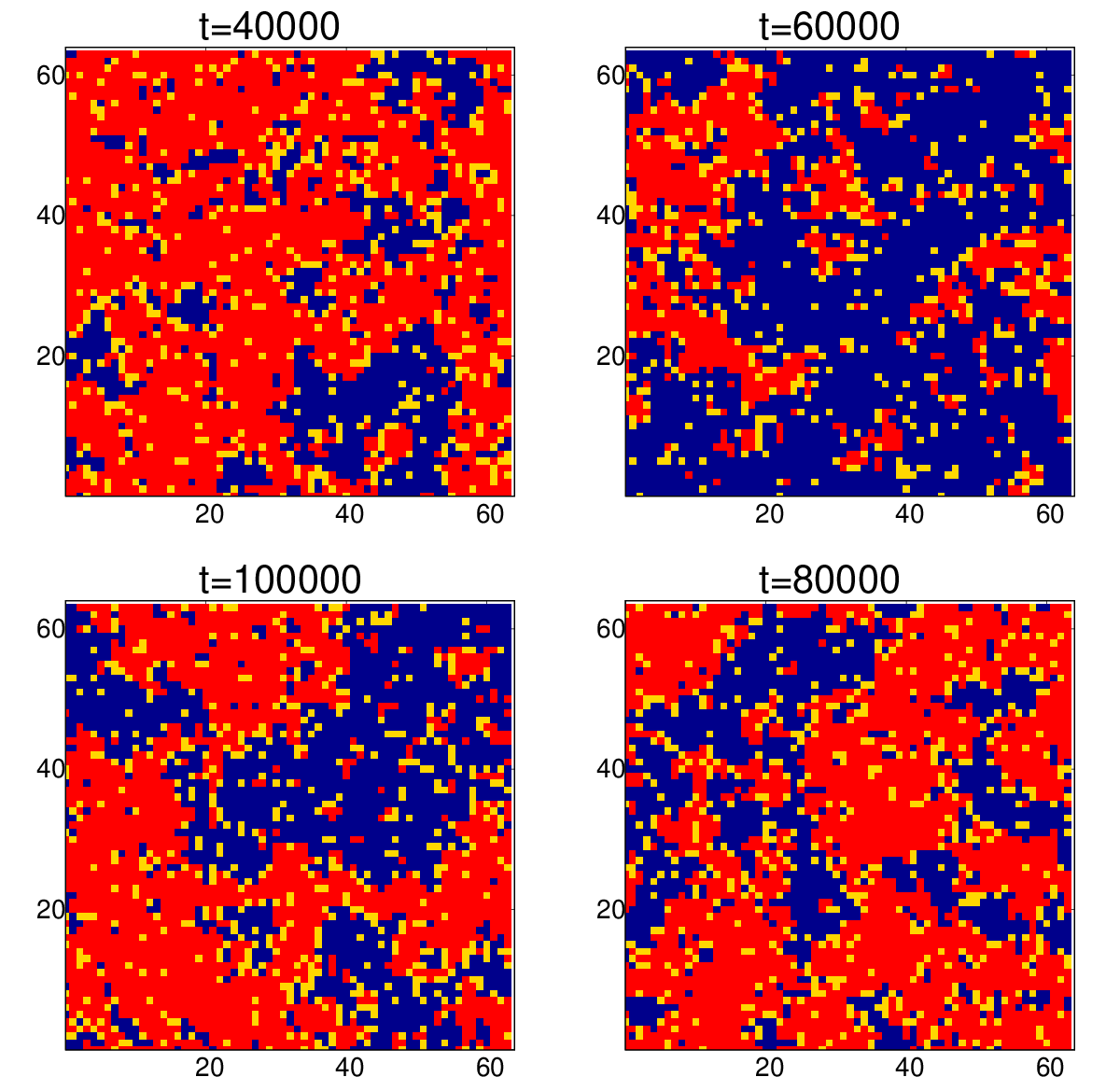}
    \caption{Snapshots of the 2$d$ lattice space at different time step 
 above the phase boundary. Red, blue, and yellow dots denote $\pm 1$ and $0$ states respectively. 
    \label{snapshot_above_pc}}
\end{figure}

On the other hand, the data collapse observed for $p>p_T$ yields 
$\alpha=0.5$ (as shown in Fig. \ref{collapse2}), which is consistent with the behavior expected in an unbiased random walk. The Gaussian function is written as $F(z)\propto e^{-\frac{z^2}{2\sigma^2}}$, where $z=X/t^{\frac{1}{2}}$. The distribution width
$\sigma$, is observed  to diverge as $|p-p_T|^{-\delta}$, plotted in  Fig. \ref{width} for different $q$ values. $\delta$  shows no systematic dependence on the exact location on the phase boundary where it is calculated and lies within a small range;  $\delta \approx 0.84$.
Snapshots in the disordered phase taken at different times show that the states of the spins are changing considerably over time which will show
typical oscillations of the order parameter about zero for a single configuration
(Fig. \ref{snapshot_above_pc}). It is interesting that both below and above the phase boundary, zero opinions are much less in number 
in comparison to the others.


\subsection{Exploring phase transitions by finite size scaling method}

Up to this point, we have discussed the results of the virtual walk generated from the BChS model with extreme switches. In this section, we  report the results for the critical points in the $p-q$ space obtained directly from the simulation of the BChS model. The most prevalent approach for investigating order-disorder phase transitions involves the finite-size scaling of the quantities like Binder cumulant, the order parameter etc.  In this particular model,  the critical points can be expressed as ($p_c, q_c$).
The order parameter for the system is the average of all opinions: $ O  = \frac{1}{N} \sum_i{o_i}$ and the 
 fourth order Binder cumulant is   $U= 1 - \frac{\langle O^4 \rangle}{3\langle O^2 \rangle^2}$. The angular brackets indicate the ensemble average.

We conducted Monte Carlo simulations for various system sizes $L$ varying between 12 and 64. The simulations were run with a sufficient number of time steps to allow measurable quantities to reach a steady value. Subsequently, we calculated the ensemble averages of these values. The number of configurations ranged from 2000 to 1000 with an increase in system size. It is well known that the scaling behavior of the Binder cumulant and the order parameter are as follows:   $U = f_1((x-x_c)L^{\frac{1}{\nu}})$ and $\langle |O| \rangle = L^{-\frac{\beta}{\nu}}f_2((x-x_c)L^{\frac{1}{\nu}})$, where the phase transition is driven by the parameter $x$ and occurs at $x=x_c$. 
$\nu$ and $\beta$ are critical exponents associated with the correlation length and order parameter respectively. 

\begin{figure}[h]
\centering
\includegraphics[width=8.5 cm]{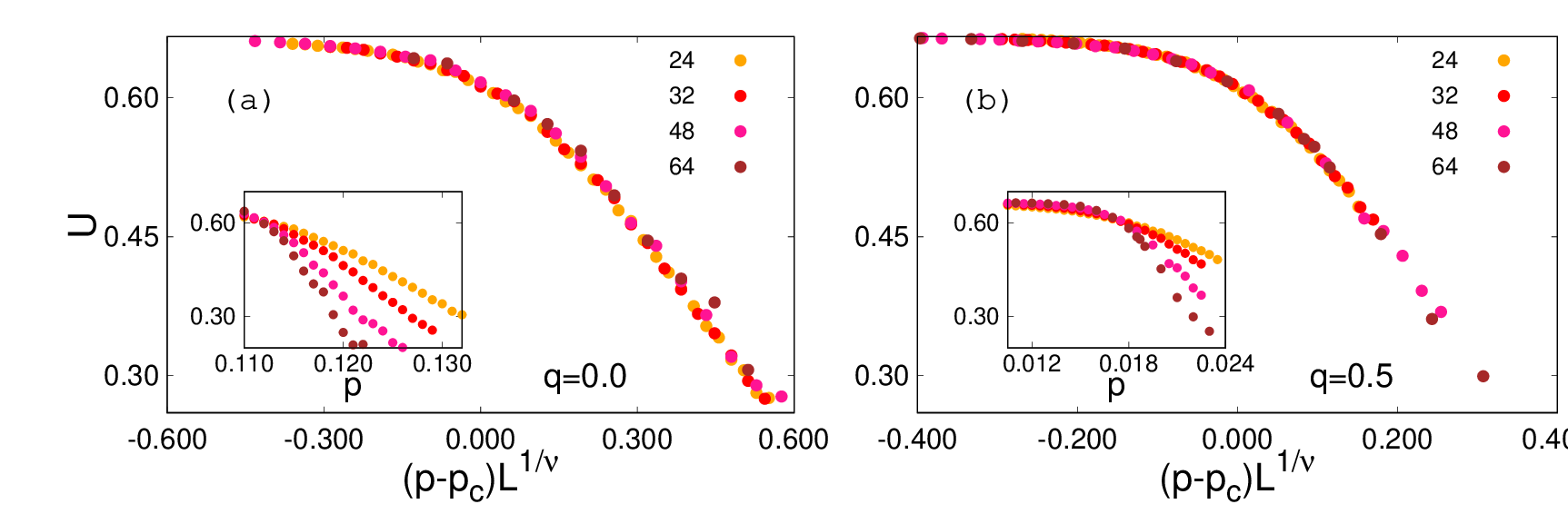}
\caption{ The finite-size scaling behavior of the Binder cumulant $U$ as a function of $p$  is illustrated for two cases: (a) $q=0.0$  and (b) $q=0.5$. Notably, data collapses are observed with critical values of $p_c=0.1110$ and $p_c=0.0170$ in (a) and (b) respectively. In both cases, the critical exponent $\nu$ is estimated to be approximately $1.0 \pm 0.05$. The inset displays the unscaled raw data for $U$ versus $p$.}
\label{fss1}
\end{figure}

\begin{figure}[h]
\centering
\includegraphics[width=8.5 cm]{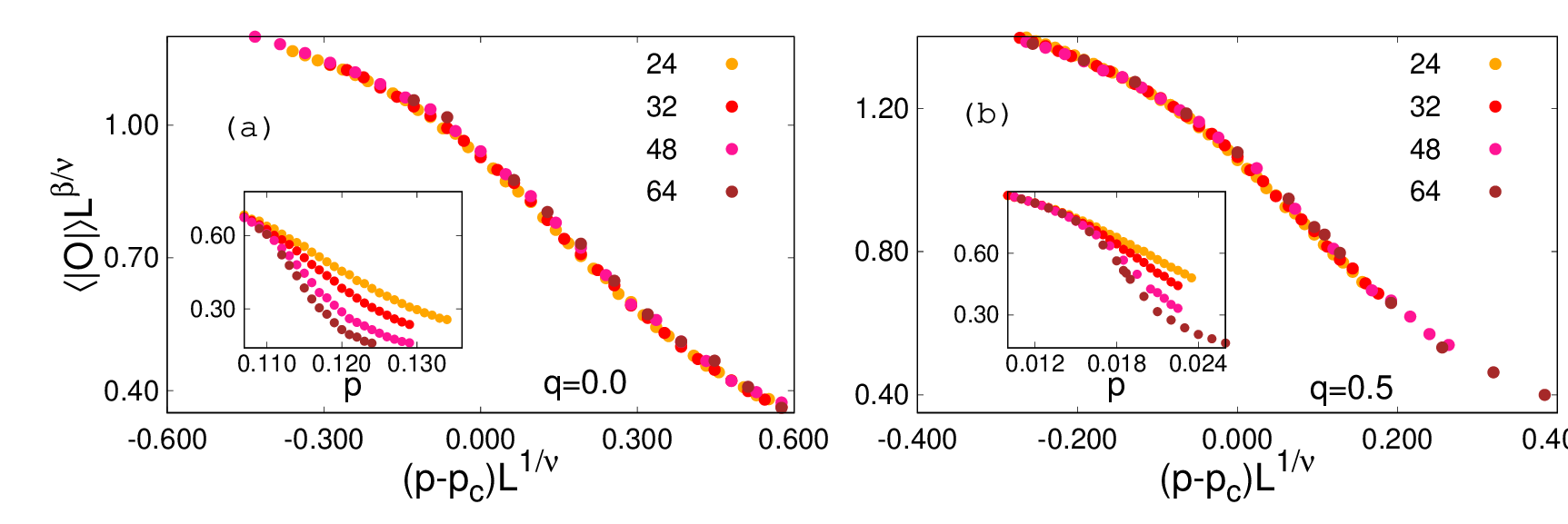}
\caption{
The data collapse of the scaled order parameter $\langle |O|\rangle$ for different system sizes is shown for two cases: (a) $q=0.0$ and (b) $q=0.5$ using the critical values of $p_c$ and $\nu$ obtained from the analysis of the Binder cumulant (Fig. \ref{fss1}). In both instances, the critical exponent $\beta$ is estimated to be  $0.125 \pm 0.001$. The inset displays the unscaled raw data $\langle O \rangle$ vs $p$.}
\label{fss2}

\end{figure}

Our aim is to check whether ($p_T,q_T$) and ($p_c,q_c$) are close enough so that it can be concluded that the virtual walks bear the signature of the phase transition. Here we have compared the values for three particular $q$ values.
Fixing the value of $q$ as $q_c$, we determined $p_c$ by identifying the crossing points of the Binder cumulant for different system sizes. Additionally, by employing the data collapse technique, we estimated the critical exponent $\nu$ to be very close to 1 (see Fig. \ref{fss1}). Using this value of $\nu$, we estimated the critical exponent $\beta \approx 0.125 $ from the data collapse of the scaled order parameter shown in Fig. \ref{fss2}. Both these exponent values are very close to the exact values known for the 
2$d$ Ising model. 

For $q=0.0$ and $0.5$, we obtain $p_c\approx0.1110$ and $\approx0.0170$ respectively. From the walk picture,  the corresponding threshold values are $p_T\approx0.1150$ and $\approx0.0185$ respectively, which are fairly close to the results obtained using finite size scaling.

\begin{figure}[h]
    \centering
    \includegraphics[width=8.5 cm]{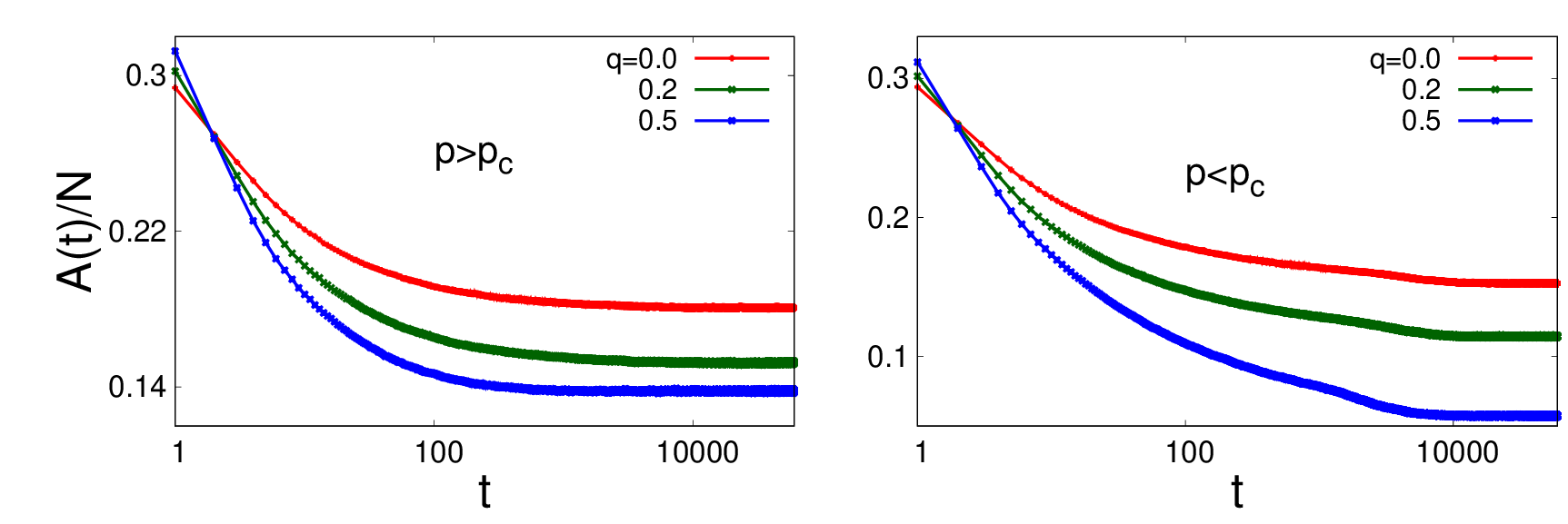}
    \caption{The variation of  the fraction of the number of active agents $A$ among the total number of agents $N$ is plotted against  time, for different values of $q$, above and below the phase boundary.  
    \label{active_agents}}
\end{figure}

\section{Discussions and Conclusions}
  In the present work, we have obtained a phase diagram in a two-parameter kinetic exchange model of opinion dynamics with three opinion states. The two parameters
  represent the fraction of negative interactions and the probability of extreme switches of opinions. By simulating the agent-based model on a 2$d$ 
  lattice, we generated the walks corresponding to each agent's opinion state in a 1$d$ virtual space. 
  Analysis of the distribution $S(X,t)$ of the displacements shows a change in the nature of the walk above threshold values of the parameters. Below these values, 
  the distribution is double peaked and the data for different times can be approximately collapsed using a scaling variable $X/t$ which reveals the nearly ballistic nature of the 
  walk. The nearly ballistic nature and the double peaked structure with peaks occurring at nonzero values of $X$ indicate that most of the opinions continue in a state of either $+1$ or $-1$. This is then a   partially
  ordered state. It may be mentioned here that for the zero opinion, no displacement is occurring in the walk and it  does not significantly affect the walk's nature. Above the threshold values, we get Gaussian distributions centered at zero with the scaling variable $X/t^{1/2}$ indicating a diffusive walk.  This means that the opinions are changing continuously and randomly in time and the system is disordered. 
  Hence the two regions below and above the threshold values are identified as ordered and disordered regions as had been observed earlier for such virtual walks corresponding
  to other dynamical systems. 

  To establish that indeed the walk changes its nature at critical points, we have also located the critical points for three $q$ values, using finite size scaling. The results 
  agree fairly well, the small discrepancy may be due to finite size effects. 
   The critical exponents $\nu$ and $\beta$ are estimated and found to be close to the Ising exponents in two dimensions, observed already for the one-parameter model without 
  extreme switches \cite{sudip}. One can also compare the results with the mean-field case \cite{kbps3},  where the phase boundary was obtained as 
  a straight line. Except for $q=1$, the mean field phase boundary lies above that of the 2$d$ one which is a logical result. In both the mean-field and two-dimensional versions, the role of $q$ is to provide additional noise without changing the universality class. Interestingly, in both cases, we find that the system becomes totally disordered when the probability of extreme switches is unity, $i.e.$ when $q=1$. In the mean-field case, it was shown that for $q=1$, the model becomes identical to a voter model with binary opinion values, we conjecture that the same happens for the finite-dimensional case as essentially the presence of $q$ decreases the probability of having a zero opinion. This is aptly reflected in the snapshots for a nonzero value of $q$.  

  From the walk picture, a critical exponent $\delta$ related to the diverging  width of the Gaussian distribution is obtained.  Like the static critical exponents  $\delta$, characterising the distribution of the walk in the disordered phase, is also $q$ independent. 
  Hence the  signature of the phase transition is contained in the distribution in this manner as well.
  As $\delta$ apparently  
   cannot be related to any static critical exponent,
  we claim it is  a new exponent. 

  We also obtained the result that the ordered phase is not an absorbing one in general from the results obtained so far. It  is known that for $p=0, q=0$, 
  an absorbing state can be  reached  but may take a very long time \cite{brexit}.
The present study indicates that a  detailed study of the dynamics including $q$ and the question of reaching an absorbing state   
could be interesting.  for the case $p=q=0$, the presence of zero opinion states at the boundary of $+1$ and ${-1}$ domains was responsible for the slow dynamics.  With  a nonzero value of $q$, the population of zero opinions 
decreases and such states could be  less significant for the  dynamics.
A preliminary study of the active bonds show that indeed it decreases $q$ is made larger in both the disordered and ordered states (Fig. \ref{active_agents}). That the active bond fraction remains non-zero at long times also supports the fact that both the phases are active.   
A detailed study of this and dynamics in general  could be interesting, however, 
at the moment  we restrict to presenting the walk features and the resulting phase diagram only. 

  One more point is there to be noted. In some earlier studies of the virtual walks, a crossover behavior in time had been observed in the ordered phase. Here, however, no such tendency is noted. 

  In conclusion, we find that a phase diagram for the two dimensional BChS model with extreme switches can be obtained using the walk picture. A usual finite size analysis has also been done to show that the criticality is of Ising class.  One more exponent entirely related to the virtual walks has been  obtained from the divergence of the width of the distribution above the critical points. For the two dimensional voter model, a similar divergence was found \cite{pratik}, however, for the mean field BChS model,  the width was found to be independent of $p$ above $p_c$ \cite{surojit1}. This feature therefore needs to investigated more by studying other models.

\begin{acknowledgments}
PS acknowledges financial support from CSIR scheme 
03/1495/23/EMR-II.
\end{acknowledgments}

\appendix



\end{document}